\documentstyle[12pt]{article}
\begin{document}
\title{High-mass diffraction in the QCD dipole picture }
\author{ A.Bialas\thanks{ On leave from Institute of Physics, Jagellonian
University, Reymonta 4,  30-059 Cracow, Poland}, H.Navelet and R.Peschanski  \\CEA, Service de Physique
Theorique, CE-Saclay\\ F-91191 Gif-sur-Yvette Cedex, France}
\maketitle
\begin{abstract}
Using the QCD dipole picture of the BFKL pomeron,  the cross-section of single diffractive dissociation of virtual photons at high energy and large diffractively excited masses is calculated. The calculation takes into account the full impact-parameter phase-space and thus allows to obtain an exact value of the triple BFKL Pomeron vertex. It appears large enough to compensate the perturbative 6-gluon coupling factor $\left(\frac {\alpha}{\pi}\right)^3$ thus suggesting a rather appreciable diffractive cross-section.
\end{abstract}
{\bf 1.} Among the resummation properties of the perturbative expansion of QCD, it has been shown that QCD predictions for onium-onium scattering  at high energy can be obtained using the QCD dipole picture \cite {mu1,mu2,mu3}, equivalent to BFKL dynamics \cite {li1}. In these processes, an onium is considered to be a $q\!-\!\bar q$ state of small transverse diameter (i.e. large mass scale)
$r_0 \approx 2/Q_0.$ In the same framework, deep inelastic $\gamma^*$-onium scattering can be calculated considering a fluctuation of $\gamma^*$  into 
$q\!-\!\bar q$ components of transverse diameter $\bar r \approx 2/Q$ and its interaction with the onium through BFKL dynamics. Interestingly enough, $q\!-\!\bar q$ -onium scattering with two a-priori different scales $r_0$ and $\bar r$ leads to a pattern of scaling violations of the onium structure functions at small $x_{Bj}$ similar \cite {na1} to those observed with the proton at HERA.

Indeed, it is known since some time  that the data on total
$\gamma^*-p$ cross-section \cite{he1}
can be successfully described \cite{na1} by the formula derived
\cite{na1,bi1} in the QCD dipole picture
\begin{eqnarray}
F_2(x_{Bj},Q^2)= \frac{11\pi}{64} N \alpha^2 e_f^2 n_{eff}\  
e^{\Delta_PY}
\left(\frac{2a(Y)}{\pi}\right)^{\frac12}\frac{Qr_0}{2}
e^{-\frac{a(Y)}{2}\ln^2 (r_0Q/2)}                 \label{i1}
\end{eqnarray}
where
\begin{equation}
Y=\ln\left(\frac{c}{x_{Bj}}\right)      \label{ii1}
\end{equation}
and $c$ is a constant fixing the scale of rapidity in the process. Furthermore,
\begin{equation}
a(Y) = \frac{\pi}{7\alpha N \zeta(3) Y}, \ \ \     1+\Delta_P=1+ \frac{4\alpha N  } {\pi} \ln 2  \label{iv1}
\end{equation}
is the BFKL pomeron intercept, $\alpha$ is the strong coupling constant
(treated as a fixed number),
$N=3$ is the number of colours and $e_f^2$ is the sum of the squares of
charges of the quarks     participating in the process. $n_{eff}$ is the
effective number of  dipoles in the proton and $r_0$ is their average
transverse diameter. Eq.(\ref{i1}) contains 4 unknown parameters: $\Delta_P,c,
n_{eff}$ and $r_0$. Three of them can be fixed by a fit to the data
\cite{na1,ro1}\footnote{The presently accepted values are \cite{ro1}:
$\Delta_P= .282,\;\;\; c=1.75,\;\;\; n_{eff}e^2_f = 3.8,\;\;\; Q_0\equiv
\frac2{r_0} = .522$ GeV.}.

{\bf 2.} This success of the dipole picture invites one to apply it to other
diffractive processes. The high-mass diffractive dissociation of virtual
photons is both theoretically \cite{mu2,bi2,ot1} and experimentally \cite{he2}
an attractive case to study. In a recent paper by two of us \cite{bi2}  a formula has been derived for this cross-section in the limit of large impact parameters. This allowed to study
its general properties but, unfortunately, could not give a well justified
normalization of the integrated cross-section which is being measured in
experiment. Indeed, it was shown already for the total $\gamma^*$-onium
cross-section that the region of impact parameter
comparable to that of the two colliding onia  plays an important
role in the process and  must be treated with a great care \cite{na2}. We shall
see below that the same is true for the diffractive process we consider.

In the present investigation we calculate the  cross-section directly, without relying on the large
impact parameter approximation. For the  diffractive structure function  we obtain
\begin{eqnarray}
\frac {dF^D_{T,L}}{d^2k} = \frac{Q^2}{4\pi^2\alpha_{em}}
x_P^{-1}\frac{d\sigma(x_{Bj},x_P,Q^2,k)}{dy d^2k}\ = 
\nonumber \\
 \frac{\alpha^5 N^2e_f^2}{8\pi^7}n_{eff}^2\ e^{2\Delta_P y} x_P^{-1}
\left(\frac{2a(y)}{2\pi}\right)^3 \nonumber \\  \ \times       
r^2_{eff}(k)e^{-a(y) \ln^2(kr_{eff}(k)/2)} \int\frac{d\bar{\gamma}}{2\pi i}
\left(\frac{Q}k\right)^{2-\bar{\gamma}}
 \beta^{-\Delta(\bar{\gamma})}\ 
 g_{3\cal P}(\bar{\gamma})S_{T,L}(\bar{\gamma}) ,                   \label{vi1}
\end{eqnarray}
where
\begin{equation}
\Delta({\gamma}) = \frac {\alpha N}{\pi} \left\{2\psi(1)-\psi (1-\gamma/2)
-\psi (\gamma/2)\right\}\ ,
\label{avii1}
\end{equation}
\begin{equation}
S_L(\gamma)=\frac{2\Gamma^4(1+\gamma/2)\Gamma^2(2-\gamma/2)}
{\Gamma(4-\gamma)\Gamma(2+\gamma)}  \frac{\Gamma(1-\gamma/2)}{\Gamma(\gamma/2)}, \;\;\;
\frac {S_T}{S_L}=\frac{(1+\gamma/2)(2-\gamma/2)}{\gamma(1-\gamma/2)}
\label{vii1}
\end{equation}
describe the coupling to the photon and
\begin{equation}
y=\ln\left(\frac{c}{x_P}\right) ,\;\;\;
\frac{x_{Bj}}{x_P}\equiv \beta  =\frac{Q^2}{M^2+Q^2}\ .  \label{vii11}
\end{equation}
 $g_{3\cal P}$ is  the multidimensional integral related to the triple
BFKL Pomeron interaction \cite{pe1}:
\begin{equation}
g_{3\cal P}(\bar{\gamma})= \int \frac{d^2x}{\mid x_+\mid \mid x_-\mid}
\int \frac{d^2s}{\mid s_+\mid \mid s_-\mid}
 \int \frac{d^2s'}{\mid s_+'\mid \mid s_-'\mid} \mid s_--s'_-
\mid^{\bar{\gamma}-2}
\label{x1}
\end{equation}
with
\begin{equation}
x_{\pm} = x\pm\frac{n}2, \;\;\; s_{\pm}= s\pm \frac{x_+}2,\;\;\;
s_{\pm}'= s'\pm \frac{x_-}2.        \label{xi1}
\end{equation}
$n$ being an arbitrary unit vector.

$r_{eff}(k)$ 
  describes the Pomeron coupling to the target,
\begin{equation}
r_{eff}(k) = \int \ r d^2r d^2s e^{i k\cdot s} \ \Phi(r,s)\ 
W(k,r,1),
\label{axii1}
\end{equation}
where
$\Phi(r,s)$ is the normalized distribution of transverse sizes $r$ and positions $s$ of the dipoles in the target. Finally, 
\begin{equation}
W(k,x,\gamma)=\int d^2s e^{-iks}
(\mid s+\frac{x}2 \mid \mid    s-\frac{x}2\mid )^{-\gamma}
   \label{vii}
\end{equation}
is the Fourier transform of the conformal eigenvectors, which give the exact solution of the BFKL equation in impact-parameter-space \cite{li2}.

In the interesting triple-pomeron limit $\beta\rightarrow 0$ one can evaluate
the path integral in (\ref{vi1}) by the saddle point method (at $\bar \gamma
\approx 1$) with the result
\begin{eqnarray}
\frac {dF^D_{T,L}}{d^2k} =
 h_{T,L}\ \frac{\alpha^5 N^2}{2^{11}\pi^4}e_f^2\ n_{eff}^2 \ r^2_{eff}(k)\ e^{-a(y) \ln^2(kr_{eff}(k)/2)}  \ g_{3\cal P}(1)   \nonumber \\
\frac{Q}k
e^{2\Delta_P y}x_P^{-1}
\left(\frac{2a(y)}{\pi}\right)^3     \left(\frac{2a(Y\!-\!y)}{\pi}\right)^{\frac12}\beta^{-\Delta_P}
e^{-\frac{a(Y\!-\!y)}{2}\ln^2  (Q/k)} 
\label{viii}
\end{eqnarray}
where $h_T=9/2$, $h_L=1$. The triple-Pomeron integral $g_{3\cal P}(1)$ has been  calculated either analytically by two different methods
\cite{bi3,ko1} or by numerical Monte-Carlo integration \cite {bb1}. The analytical methods   yield
\begin{eqnarray}
&&g_{3\cal P}(1) \equiv 16 \pi^5\   _4F_3(\frac12,\frac12,\frac12,\frac12;1,1,1;1) \times \nonumber \\
&&\times \frac {\partial}{\partial\epsilon} {\frac {\Gamma\left(\frac12-\epsilon\right)} {\Gamma\left(1-\epsilon\right)} } \ _4F_3(\frac12-\epsilon,\frac12,\frac12,\frac12;1-\epsilon,1,1;1)\vert_{\epsilon=0} \approx \ 7766
\label{xiv1}
\end{eqnarray}
where  $_4F_3$ is the generalized hypergeometric function.

{\bf 3.} 
Let us now outline the derivation of Eqn.(\ref{vi1}). For clarity
we restrict ourselves to the case of a simple dipole target of transverse diameter $r_0.$ The generalization to a general target (onium, proton, etc...)
can be done on the same lines, as indicated at the end of this section.

Following \cite{mu2}, one writes
\begin{equation}
\frac{d\sigma}{dy d^2k} =(2\pi)^{-2}  \ \tilde{F}_d(\bar{r},r_0,Y,y,k),
\label{bi}
\end{equation}
with
\begin{eqnarray}
\tilde{F}_d(\bar{r},r_0,Y,y,k)=\int \frac{dx}{x} \frac{dx'}{x'}
\frac{d\bar{x}}{\bar{x}}\frac{d\bar{x}'}{\bar{x}'}\tau(x,\bar{x})\tau(x',\bar{x}')
 d^2s d^2s'\nonumber \\
\tilde{n}_1(r_0,x,y*,k) \tilde{n}_1(r_0,x',y*,-k)\tilde{n}_2(\bar{r},
\bar{x},\bar{x}',Y-y*,y-y*,k)\ ,   \label{ib}
\end{eqnarray}
where $\tilde n_1$ and $\tilde n_2$ are single and double dipole densities in the colliding
dipoles with rapidities $y*$ and $Y\!-\!y*$ respectively (note that the final result is independent of $y*$). $\tau$ is the two-gluon exchange dipole-dipole amplitude.

The equation for $\tilde{n}_2$ is given in \cite{mu2} (Eq.52).
It can be solved by the methods of Refs. \cite{mu3,bi3}
with the result
\begin{eqnarray}
\tilde{n}_2(\bar{r},\bar{x},\bar{x}',Y,y,k) = \frac{\alpha N}{\pi^2}
\int \frac{d\bar{\gamma}}{2\pi i} \bar{r}^{\bar{\gamma}} e^{\Delta(\bar{\gamma})(Y-y)}
\nonumber \\
\int \frac{dx_{01}}{x_{01}} x_{01}^{-\bar{\gamma}}
 \int\frac{d^2x_2 x_{01}^2}{x_{12}^2x_{02}^2}  e^{ikx_{01}/2}
\tilde{n}_1(x_{02},\bar{x},y,k) \tilde{n}_1(x_{12},\bar{x}',y,k)
\label{iia}
\end{eqnarray}
where the complex integral over $\bar{\gamma}$ goes from $1-i\infty$ to
$1+i\infty$.

To calculate the integrals over $x,x',\bar{x},\bar{x}',s,s'$ we use twice the
identity \cite {na3}
\begin{equation}
\tilde{F}^{(1)}(x_{01}, x_{01}',Y,k) =  \int\frac{dx}{x} \frac{dx'}{x'}
\tau(x,x')  n_1(x_{01},x,Y-y,k)
n_1(x_{01}',x',y,k)     \label{iii}
\end{equation}
where $\tilde{F}^{(1)}(x_{01}, x_{01}',Y,k)$ is the  dipole-dipole amplitude at fixed momentum transfer.
 We thus obtain
\begin{eqnarray}
\tilde{F}_d(\bar{r},r_0,Y,y,k)=
&& \frac{\alpha N}{\pi^2} \int
\frac{d\bar{\gamma}}{2\pi i}\bar{r}^{\bar{\gamma}} e^{\frac{\alpha N}{\pi}
\chi(\bar{\gamma})(Y-y)}\frac1{2\pi} \int\frac{d^2x_{01}} {x_{01}^2}
x_{01}^{-\bar{\gamma}} e^{ikx_{01}/2}
\nonumber \\ &&\int\frac{d^2x_2 x_{01}^2}{x_{12}^2x_{02}^2}
\tilde{F}^{(1)}(r_0,x_{02},y,k)\tilde{F}^{(1)}(r_0,x_{12},y,-k)   \label{iv}
\end{eqnarray}
where \cite{na2}:
\begin{eqnarray}
\tilde{F}^{(1)}(r,r',y,k)= -\frac{\alpha^2}{\pi}rr'\int\frac{d\gamma}{2\pi i}
e^{\Delta(\gamma)y}\left(\frac{r}{r'}\right)^{\gamma-1}(1-\gamma)^2h(\gamma)
\nonumber \\ W(k,r,\gamma)W(k,r',2-\gamma)      \label{v}
\end{eqnarray}
with
\begin{equation}
h(\gamma)=\frac4{(2-\gamma)^2\gamma^2}.         \label{vi}
\end{equation}

Here it is worth to stress that the formula (\ref{v}) with $W(k,x,\gamma)$
given by (\ref{vii}) is
{\it
exact} (within the approximations inherent in the dipole approach). It is at this
point that we deviate from the  treatment of Ref.\cite{bi2} where, following
\cite{mu3} the factor
$(\mid s+\frac{x}2 \mid \mid    s-\frac{x}2\mid )^{-\gamma}$ in (\ref{vii})
was replaced by its asymptotic form $\mid s \mid^{-2\gamma}$. This
simplification allowed a first estimate of the solution but, while removing   the singularities at $s=\pm\frac{x}2,$ it changed
 the absolute value of the integral.

 Substituting
(\ref{v}) into
(\ref{iv}) we obtain
\begin{eqnarray}
\tilde{F}_d(\bar{r},r_0,Y,y,k)=
 \frac{\alpha^5 N}{\pi^4}(r_0)^4
 \int\frac{d\bar{\gamma}}{2\pi i}\bar{r}^{\bar{\gamma}}
 e^{\Delta(\bar{\gamma})(Y-y)}  \nonumber \\
 \int\frac{d\gamma}{2\pi i}e^{\Delta(\gamma)y} W(k,r_0,2-\gamma) r_0^{-\gamma}
(1\!-\!\gamma)^2h(\gamma)                       \nonumber \\
\int\frac{d\gamma'}{2\pi
i}e^{\Delta(\gamma')y}W(-k,r_0,2\!-\!\gamma')r_0^{-\gamma'}
(1\!-\!\gamma')^2h(\gamma') Z(k,\bar{\gamma},\gamma,\gamma')  \label{viia}
\end{eqnarray}
where
\begin{eqnarray}
Z(k,\bar{\gamma},\gamma,\gamma')=
\frac1{2\pi} \int\frac{d^2x_{01}} {x_{01}^2}
x_{01}^{-\bar{\gamma}} e^{ikx_{01}/2}
 \int d^2x_2\ \frac{ x_{01}^2}{x_{12}^2x_{02}^2} \nonumber \\
x_{02}^{\gamma}W(k,x_{02},\gamma)
x_{12}^{\gamma'} W(-k,x_{12},\gamma').  \label{viiiZ}
\end{eqnarray}
Using now (\ref{vii}) one can perform the integration over $d^2x_{01}$ and
sort out the dependence of $Z$ on $k.$ One obtains
\begin{eqnarray}
Z(k,\bar{\gamma},\gamma,\gamma')=
2^{3-\bar{\gamma}-\gamma-\gamma'} k^{\bar{\gamma}+\gamma+\gamma'-4}
\frac{\Gamma(2-\frac{\bar{\gamma}+\gamma+\gamma'}2)}
{\Gamma(\!-\!1+\frac{\bar{\gamma}+\gamma+\gamma'}2)}\ 
g_{3\cal P}(\bar{\gamma},\gamma,\gamma')  \label{xiv}
\end{eqnarray}
where (c.f. definition (\ref {xi1}))
\begin{eqnarray}
g_{3\cal P}(\bar{\gamma},\gamma,\gamma')  =
 \int \frac{d^2x}{\mid x_+\mid^{2\!-\!\gamma} \mid
x_-\mid^{2-\gamma'}} \ \ \ \ \nonumber \\ 
\int \frac{d^2s}{\left(\mid s_+\mid \mid s_-\mid\right)^{\gamma}}
 \int \frac{d^2s'}{\left(\mid s_+'\mid \mid s_-'\mid\right)^{\gamma'}}
 \mid s_--s'_- \mid^{\bar{\gamma}+\gamma+\gamma'-4}.
\label{xv}
\end{eqnarray}
It is not difficult to modify this result to account for targets composed of  dipoles. One simply has to integrate over the distribution of the dipoles in the target. Consequently Eqn. (\ref {viia}) becomes:
\begin{eqnarray}
\tilde{F}_d(\bar{r},Y,y,k)=
 \frac{8\alpha^5 N}{\pi^4}\ k^{-4}\ 
 \int\frac{d\bar{\gamma}}{2\pi i}\left(\frac {k \bar{r}}2\right)^{\bar{\gamma}}
 e^{\Delta(\bar{\gamma})(Y-y)}  \nonumber \\
 \int\frac{d\gamma}{2\pi i}e^{\Delta(\gamma)y} \left(\frac {k r_{eff}(k,\gamma)}2\right)^{\gamma}
(1-\gamma)^2h(\gamma)     \ \ \ \ \ \                   \nonumber \\
\int\frac{d\gamma'}{2\pi
i}e^{\Delta(\gamma')y}\ \left(\frac {k r_{eff}(k,\gamma')}2\right) ^{\gamma'}(1-\gamma')^2h(\gamma') \frac{\Gamma(2-\frac{\bar{\gamma}+\gamma+\gamma'}2)}
{\Gamma(-1+\frac{\bar{\gamma}+\gamma+\gamma'}2)}\ 
g_{3\cal P}(\bar{\gamma},\gamma,\gamma')
\label{xiav}
\end{eqnarray}
where
\begin{equation}
r^{\gamma}_{eff}(k,\gamma) = \int \ r^{2-\gamma} d^2r d^2s e^{i k\cdot s} \ \Phi(r,s)\ 
W(k,r,2-\gamma).
 \label{xiaa}
\end{equation}

Since  the formula (\ref{xiav}) is anyway valid only in the limit $x_p
\rightarrow
0$, the next natural step is to evaluate the integrals over
$\gamma$ and
$\gamma'$ by the saddle point method.
When  the result is
sandwiched between the photon wave functions \cite {bj2} and integrals over $\bar{r}$ and
quark light-cone fractions performed, one obtains (\ref{vi1}).

{\bf 4.} In conclusion, using the technique of the QCD dipole picture, we have evaluated the BFKL Pomeron contribution to high-mass diffraction dissociation of the virtual photon. The result is a compact formula which - we feel - may be a good starting point for a phenomenological discussion of the photon diffractive dissociation on a proton target in a QCD framework.
 
Several comments are in order. 

({\bf i})
Eqn. (\ref{vi1}) was obtained by the evaluation of the inverse Mellin transforms in (\ref {xiav}) by the saddle-point method. This is expected to be a reasonable approximation as long as $x_P$ is small enough. One often finds convenient, however, to have a formula for $dF^D_{T,L}/d^2k$
which exhibits explicitely the path integrals in $\gamma$ and $\gamma'.$
The result reads
\begin{eqnarray}
\frac {dF^D_{T,L}}{d^2k} =
 \frac{\alpha^5 N^2 e^2_f n^2_{eff}}{2\pi^9}\ k^{-2}\ x^{-1}_P\ 
 \int\frac{d\bar{\gamma}}{2\pi i}\left(\frac {Q}k\right)^{2-\bar{\gamma}}
 e^{\Delta(\bar{\gamma})(Y-y)} S_{T,L} (\bar{\gamma}) \nonumber \\
 \int\frac{d\gamma}{2\pi i}e^{\Delta(\gamma)y} \left(\frac {k r_{eff}(k,\gamma)}2\right)^{\gamma}
(1-\gamma)^2h(\gamma)                       \nonumber \\
\int\frac{d\gamma'}{2\pi
i}e^{\Delta(\gamma')y}\ \left(\frac {k r_{eff}(k,\gamma')}2\right)^{\gamma'} (1-\gamma')^2h(\gamma') \nonumber \\ \frac{\Gamma(2-\frac{\bar{\gamma}+\gamma+\gamma'}2)}
{\Gamma(-1+\frac{\bar{\gamma}+\gamma+\gamma'}2)}\ 
g_{3\cal P}(\bar{\gamma},\gamma,\gamma')\ \ \ \ 
\label{xicom}
\end{eqnarray}

({\bf ii})
It is illuminating to compare these results  with
the one obtained by integration of the asymptotic formula in impact parameter
\cite{bi2}, viz.
\begin{eqnarray}
F^{Ddipole}_{Tasymptotic} =
 \frac{9\pi G \alpha^5 N^2}{8}e_f^2 n_{eff}^2    \ \ \ \ \nonumber \\
\frac{r_0Q}2
e^{2\Delta_P y}x_P^{-1}
\left(\frac{2a(y)}{\pi}\right)^3    
\left(\frac{2a(Y\!-\!y)}{\pi}\right)^{\frac12}\beta^{-\Delta_P}
e^{-\frac{a(Y-y)}{2}\ln^2  (r_0Q/2)}               \label{xv1}
\end{eqnarray}
($G\approx .915$ is Catalan's constant). Equation (\ref {xv1}) has been obtained for diffraction on a single dipole. To compare, one has to take $r_{eff}^{\gamma} = r_0^{2-\gamma} W(k,r_0,2-\gamma) $  in formula (\ref {xicom}), integrate over $d^2k$ and evaluate the path integral by the saddle-point method. The integration over $d^2 k$ is done with help of formula
\begin{equation}
\int d^2 k k^{\bar \gamma -2} W(k,r_0,1)\ W(-k,r_0,1) = \pi (r_0/2)^{-\bar \gamma} \frac {\Gamma (\bar \gamma /2)} {\Gamma (1-\bar \gamma /2)} U_2(\bar{\gamma}),
\label{viij1}
\end{equation}
where
\begin{equation}
U_2(\bar{\gamma})= \int \frac{d^2t}{\mid t_+\mid \mid t_-\mid}
 \int \frac{d^2t'}{\mid t_+'\mid \mid t_-'\mid} \mid t-t' \mid^{-\bar{\gamma}}
\label{viii1}
\end{equation}
with the two-dimensional vectors $t_{\pm}$ and $t_{\pm}'$ given by
\begin{equation}
t_{\pm}= t\pm \frac{n}2, \;\;\;\;  t_{\pm}'= t'\pm \frac{n}2.\label{ix1}
\end{equation}
 The  integral $U_2(1)$ yields \cite {bi3} :
\begin{equation}
U_2(1)= 2\pi^4\;_3F_2^2( \frac12, \frac12, \frac12;1,1;1)=\frac {\Gamma ^8(1/4)}{8\pi^2}\approx 378,
\label{xii1}
\end{equation}
where  $_3F_2$ is the generalized hypergeometric function.

The resulting formula shows two striking features:

(a) The dependence on $Q^2$, $\beta$ and $x_P$ is {\it exactly the same} as in  formula (\ref {xv1}). Thus the asymptotic formula correctly reproduces the dependence
of the structure function on the kinematic variables. It is most likely yet
another  consequence of the global conformal invariance of the theory
\cite{li2}.

(b) The normalization factors are very different, however, the exact
approach giving a much larger diffractive cross-section than the asymptotic
formula: the enhancement factor given by the ratio
\begin{equation}
\frac{F^{Ddipole}}{F^{Ddipole}_{asymptotic}} = \frac  {U_2(1) g_{3\cal P}(1)}{128\pi^4G}\approx 257
\label{xvi1}
\end{equation}
is indeed formidable. This is related to the fact that the asymptotic
formula ignores the strong enhancement in the singular region where the impact
parameter is close to the tranverse sizes of the interacting dipoles.  For other targets (e.g. a proton) this enhancement factor is  modified, however, depending on the shape of the distribution $\Phi(r,s).$

({\bf iii})
In order to compare our result with the ordinary triple Regge formalism \cite {ka1}, we evaluate the effective triple Pomeron coupling emerging from Eqn.(\ref{vi1}). For this sake, let us consider the ratio of the Mellin transforms of the diffractive and total $\gamma^*$-proton cross-sections, defined by
$\sigma \equiv \int \frac {d\bar\gamma}{2\pi i} \left(\frac {Q r_0}2\right)^{2-\bar\gamma}\tilde \sigma.$ Using  (\ref{vi1}) and Ref. \cite {bi5}, formula (34), one obtains at the saddle-point $\bar\gamma =1,$ 
\begin{equation}
\frac 1{\tilde \sigma (\beta)} \frac {d\tilde \sigma}{dy d^2k} \vert_{\bar \gamma =1} = \frac {\alpha^3N}{64 \pi^5} n_{eff} e^{2\Delta_P y} \left(\frac {2a(y)}{\pi}\right)^3 g_{3\cal P}(1) \frac {2 r_{eff}^2(k)}{kr_0} e^{-a(y)\ln ^2 \left(\frac {kr_{eff}(k)} {2}\right)}
\label{1}
\end{equation}

Eqn. (\ref {1}) reveals some  similarity with that obtained from Regge
theory in the triple Pomeron region, which for diffractive excitation  on the target $B$ gives 
\begin {equation}
\frac 1{\sigma}\frac {d\sigma}{dy d^2k} = \frac {g_B^2 (k)}{16 \pi^2} e^{2 (\alpha_P (k) -1)\ y}
 G_{3\cal P}(k^2),
\label {2}
\end {equation}
where  $g_B(k)$ is the coupling of the Pomeron to the proton and $G_{3\cal P}(k^2)$ is the triple Pomeron coupling, in the case 
when the Pomeron is a simple Regge pole \cite {ka1}. One observes, however, that Eqn.(\ref {1}) does not correspond to the exchange of Regge poles but contains additional logarithmic corrections. This is not surprising because the BFKL Pomeron corresponds to a Regge cut in the complex angular momentum \cite {li1}.

Putting aside the differences between (\ref {1}) and (\ref {2}) due to the different dynamical content, we are led to estimate an effective triple Pomeron coupling as:
\begin{equation}
G_{3\cal P}^{eff}(k) \approx \frac 1k \ g_{3\cal P}(1) \left(\frac {\alpha}{\pi}\right)^3 \frac N4 \left(\frac {2a(y)}{\pi}\right)^3.
\label {3}
\end {equation}
One sees that the factor $\frac 1k,$ which was already observed in Ref. \cite {mu2}, sets the scale of $G_{3\cal P}^{eff}(k)$ as expected in a conformal invariant theory. On the other hand, the presence of the factor $\left(\frac {2a(y)}{\pi}\right)^3$ reflects the logarithmic corrections to the Pomeron singularity and lowers the effective Pomeron intercept \cite {bi2}. The factor
$\left(\frac {\alpha}{\pi}\right)^3$ is naturally expected for the  perturbative 6-gluon coupling. The  value of $g_{3\cal P}(1)$ given in (\ref {xiv1}),
however, largely compensates the smalllness of the perturbative factor. One thus expects a fairly large diffractive cross-section in this framework.
  
({\bf iv})
It should be stressed that our analysis is valid only for $k$ different from 0.
The point $k=0$ (crucial e.g. for the calculation of nuclear shadowing \cite {bi4}) is special and thus requires a separate discussion. Note that -until now- only the forward BFKL amplitudes were subject to a stringent experimental testing
\cite{na1,ro1}. Thus our calculation enlarges the possibilities of investigating the relevance of BFKL dynamics in the as yet poorly explored regime of $k$ different from $0.$

\bigskip
 
{\bf Acknowledgements.} We would like to thank Ch.Royon for 
discussion and communicating his unpublished results. The discussions with
A.Kaidalov  are highly appreciated. AB thanks J.Zinn-Justin for
the kind hospitality at Service de Physique Th\' eorique de Saclay. This work
was supported in part by the KBN Grant No 2 P03B083 08 and
by PECO grant from the EEC Programme "Human Capital Mobility",
Network "Physics at High Energy Colliders", Contract No ERBICIPDCT 940613.
\eject

\end{document}